\documentclass[twoside,twocolumn,9pt]{article}
\usepackage{extsizes}
\usepackage[super,sort&compress,comma]{natbib} 
\usepackage[version=3]{mhchem}
\usepackage[left=1.5cm, right=1.5cm, top=1.785cm, bottom=2.0cm]{geometry}
\usepackage{balance}
\usepackage{times,mathptmx}
\usepackage{sectsty}
\usepackage{graphicx} 
\usepackage{lastpage}
\usepackage[format=plain,justification=justified,singlelinecheck=false,font={stretch=1.125,small,sf},labelfont=bf,labelsep=space]{caption}
\usepackage{float}
\usepackage{fancyhdr}
\usepackage{fnpos}
\usepackage[english]{babel}
\usepackage{array}
\usepackage{droidsans}
\usepackage{charter}
\usepackage[T1]{fontenc}
\usepackage[usenames,dvipsnames]{xcolor}
\usepackage{setspace}
\usepackage[compact]{titlesec}
\DeclareMathOperator{\Tr}{Tr}
\usepackage{amsmath, amssymb}
\usepackage{subfigure}
\usepackage{xcolor}

\usepackage{epstopdf}

\definecolor{cream}{RGB}{222,217,201}

\begin{document}

\pagestyle{fancy}
\thispagestyle{plain}
\fancypagestyle{plain}{

\fancyhead[C]{\includegraphics[width=18.5cm]{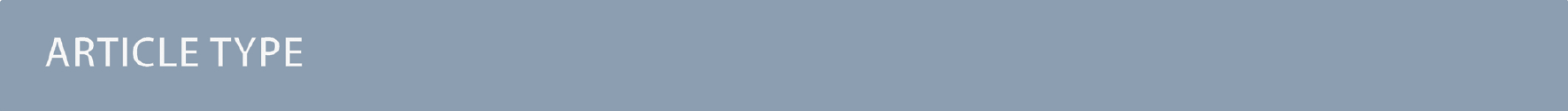}}
\fancyhead[L]{\hspace{0cm}\vspace{1.5cm}\includegraphics[height=30pt]{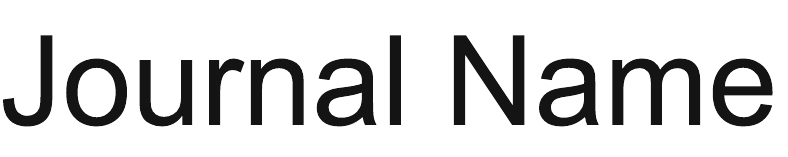}}
\fancyhead[R]{\hspace{0cm}\vspace{1.7cm}\includegraphics[height=55pt]{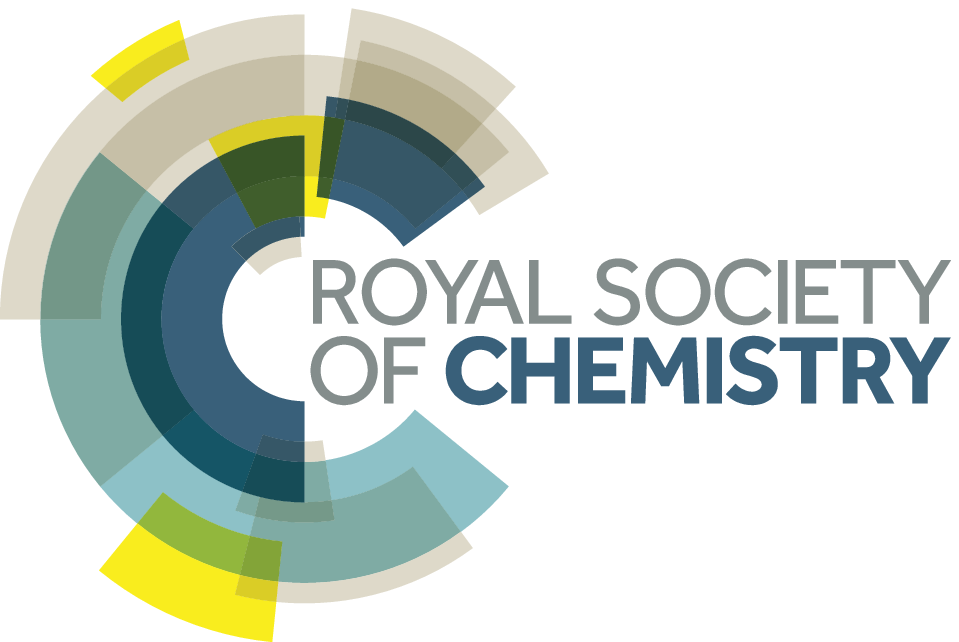}}
\renewcommand{\headrulewidth}{0pt}
}

\makeFNbottom
\makeatletter
\renewcommand\LARGE{\@setfontsize\LARGE{15pt}{17}}
\renewcommand\Large{\@setfontsize\Large{12pt}{14}}
\renewcommand\large{\@setfontsize\large{10pt}{12}}
\renewcommand\footnotesize{\@setfontsize\footnotesize{7pt}{10}}
\makeatother

\renewcommand{\thefootnote}{\fnsymbol{footnote}}
\renewcommand\footnoterule{\vspace*{1pt}%
\color{cream}\hrule width 3.5in height 0.4pt \color{black}\vspace*{5pt}} 
\setcounter{secnumdepth}{5}

\makeatletter 
\renewcommand\@biblabel[1]{#1}            
\renewcommand\@makefntext[1]%
{\noindent\makebox[0pt][r]{\@thefnmark\,}#1}
\makeatother 
\renewcommand{\figurename}{\small{Fig.}~}
\sectionfont{\sffamily\Large}
\subsectionfont{\normalsize}
\subsubsectionfont{\bf}
\setstretch{1.125} 
\setlength{\skip\footins}{0.8cm}
\setlength{\footnotesep}{0.25cm}
\setlength{\jot}{10pt}
\titlespacing*{\section}{0pt}{4pt}{4pt}
\titlespacing*{\subsection}{0pt}{15pt}{1pt}

\fancyfoot{}
\fancyfoot[LO,RE]{\vspace{-7.1pt}\includegraphics[height=9pt]{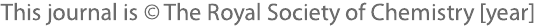}}
\fancyfoot[CO]{\vspace{-7.1pt}\hspace{13.2cm}\includegraphics{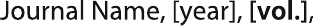}}
\fancyfoot[CE]{\vspace{-7.2pt}\hspace{-14.2cm}\includegraphics{head_foot/RF}}
\fancyfoot[RO]{\footnotesize{\sffamily{1--\pageref{LastPage} ~\textbar  \hspace{2pt}\thepage}}}
\fancyfoot[LE]{\footnotesize{\sffamily{\thepage~\textbar\hspace{3.45cm} 1--\pageref{LastPage}}}}
\fancyhead{}
\renewcommand{\headrulewidth}{0pt} 
\renewcommand{\footrulewidth}{0pt}
\setlength{\arrayrulewidth}{1pt}
\setlength{\columnsep}{6.5mm}
\setlength\bibsep{1pt}

\makeatletter 
\newlength{\figrulesep} 
\setlength{\figrulesep}{0.5\textfloatsep} 

\newcommand{\topfigrule}{\vspace*{-1pt}%
\noindent{\color{cream}\rule[-\figrulesep]{\columnwidth}{1.5pt}} }

\newcommand{\botfigrule}{\vspace*{-2pt}%
\noindent{\color{cream}\rule[\figrulesep]{\columnwidth}{1.5pt}} }

\newcommand{\dblfigrule}{\vspace*{-1pt}%
\noindent{\color{cream}\rule[-\figrulesep]{\textwidth}{1.5pt}} }

\makeatother

\twocolumn[
  \begin{@twocolumnfalse}
\vspace{3cm}
\sffamily
\begin{tabular}{m{4.5cm} p{13.5cm} }

\includegraphics{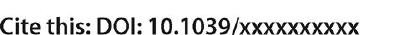} & \noindent\LARGE{\textbf{Twist--bend nematic phases of bent--shaped biaxial molecules }}\\ 
\vspace{0.3cm} & \vspace{0.3cm} \\

 & \noindent\large{Wojciech Tomczyk$^{\ast}$, Grzegorz Paj\k{a}k$^{\ast \ast}$ and Lech Longa$^{\ast \ast \ast}$} \\

\includegraphics{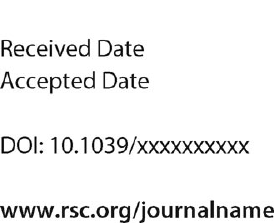} & \noindent\normalsize{How change in molecular structure can affect relative stability and structural properties of the twist--bend nematic phase (N$_\text{TB}$)? Here we extend the mean--field model\cite{ref31} for bent--shaped achiral molecules, to study the influence of arm molecular biaxiality and the value of molecule's bend angle on relative stability of N$_\text{TB}$. In particular we show that by controlling biaxiality of molecule's arms up to four ordered phases can become stable. They involve locally uniaxial and biaxial variants of N$_\text{TB}$, together with the uniaxial and the biaxial nematic phases. However, the V--~shaped molecule show stronger ability to form stable N$_\text{TB}$ than a biaxial nematic phase, where the latter phase appears in the phase diagram only for bend angles greater than $140^\circ$ and for large biaxiality of the two arms.   } \\

\end{tabular}

 \end{@twocolumnfalse} \vspace{0.6cm}

  ]

\renewcommand*\rmdefault{bch}\normalfont\upshape
\rmfamily
\section*{}
\vspace{-1cm}


\footnotetext{\textit{Marian Smoluchowski Institute of Physics, Department of Statistical Physics, Jagiellonian University, prof. S. \L ojasiewicza 11, 30-348 Krak\'{o}w, Poland.;\\ $^{\ast}$E-mail: wojciech.tomczyk@doctoral.uj.edu.pl \\$^{\ast \ast}$E-mail: grzegorz@th.if.uj.edu.pl\\ $^{\ast \ast \ast}$E-mail: lech.longa@uj.edu.pl }}





\section{Introduction}

One of the most surprising recent discovery in the field of soft matter physics is the identification of a new nematic phase, known as the nematic twist--bend phase (N$_\text{TB}$) \cite{ref1,ref2,ref3}. This phase is stabilized as a result of spontaneous chiral symmetry breaking in liquid crystalline systems composed of achiral bent--core \cite{ref4}, dimeric \cite{ref5} and trimeric \cite{ref6} mesogens. The director in N$_\text{TB}$ forms a conical helix with nanoscale periodicity, while molecular achirality implies that coexisting domains of opposite chirality are formed. Up to now, only some characteristic features of this elusive phase are known, like \emph{e.g.}, tilt angle and the pitch length or values of elastic constants in the vicinity of the uniaxial nematic (N$_\text{U}$) -- N$_\text{TB}$ phase transition \cite{ref7,ref8,ref9,ref10,ref11,ref12}. Experiments show that the N$_\text{TB}$ phase usually occurs within the stability regime of N$_\text{U}$ \cite{ref13}, but for some compounds  a direct transition between N$_\text{TB}$ and the isotropic phase (Iso) has also been found  \cite{ref14}. Although mechanism leading to long--range chiral order of N$_\text{TB}$ is to a large extent unknown, the issue of stable, modulated nematic phases has been addressed theoretically in a series of papers \cite{ref15,ref16,ref17,ref18,ref19,ref20,ref21,epje,ref22,ref23,ref24,ref25}. 

Meyer\cite{ref23,ref24} and subsequently Dozov\cite{ref25} have shown that flexopolarization can be the driving force leading to twist--bend and splay--bend distortions of the director field. Lorman and Mettout\cite{ref26,ref27}, suggested that the formation of the N$_\text{TB}$, and other unconventional periodic structures, can be facilitated by the shape of bent--core molecules, which seems to be in line with experimental results \cite{ref13, ref28} and computer simulations \cite{ref2,ref21,ref29,ref30}. 

Out of alternative theoretical approaches undertaken to tackle the nature of 
N$_\text{TB}$ we will focus on a generic, mean--field model 
introduced by 
Greco, Luckhurst and Ferrarini (GLF) \cite{ref31}. In the GLF
model N$_\text{TB}$ is treated as an inhomogeneous and
\textit{locally} uniaxial heliconical periodic distortion of the nematic phase, 
characterized at each point by the single local director
$\mathbf{\hat{n}}(\mathbf{r}) \equiv \mathbf{\hat{n}}({z})$ \cite{ref25,ref31} 
 (see Figure \ref{fig:scheme1}):
\begin{equation}\label{nn}
\mathbf{\hat{n}}(z) = [-\sin (\theta ) \sin (\phi ),\sin (\theta ) \cos (\phi ),\cos (\theta )],
\end{equation}
where $\theta$ is the conical angle and $\phi=k z=\frac{2\pi}{p} z$ 
with wave vector $\mathbf{k}=k \mathbf{\hat{z}}$ 
($k=\pm\frac{2\pi}{p}$) and 
with period $p$ 
of the phase. 
The wave vector,  being parallel to the average direction of the main director over one period $p$: $\mathbf{k} \parallel \langle \mathbf{\hat{n}} \rangle_p$, can be identified with an effective optical axis \cite{ref15,ref16,ref18}.

Heliconical precession is assumed arbitrarily to take place around the $\mathbf{\hat{z}}$--axis of the laboratory system of frame. The helix of N$_\text{TB}$ can be right--handed or left--handed, depending on the sign of $k$, and both of these mono--domains have the same free energy. Moreover, a rigid, biaxial bent--core molecule is represented by two 
mesogenic arms of cylindrical symmetry, each assumed to align preferentially to $\mathbf{\hat{n}}(z)$. The latter is taken at the position of the midpoint of the arm. Only N$_\text{TB}$, N$_\text{U}$ and the isotropic liquid can be stabilized by the GLF model.

The effective mean--field potential acting on molecular arms is defined by the well--known Maier--Saupe $P_2$ potential, with $P_2$ being the second Legendre polynomial. Despite its simplicity the GLF model correctly predicts N$_\text{U}$ 
to N$_\text{TB}$ and Iso to N$_\text{TB}$ phase transitions, weak temperature dependence of the pitch and consistent description of elastic properties of the N$_\text{TB}$. Tailoring molecules with particular shapes and interactions, it 
seems interesting to study extensions of the GLF model to molecules of more complex structure.

While there are many paths {that can be followed}, one obvious observation is that bent--shaped molecules, including the famous N$_\text{TB}$--forming compound CB7CB, can acquire biaxiality not only due to their average "V" shape, but also as a results of biaxiality of molecule's arms and conformational degrees of freedom \cite{ref32,ref33}. Importantly, they can form a stable biaxial nematic phase \cite{PhysRevE.72.051702,ref34,cm}, and, hence, this structure should be included into theoretical analysis as a possible competitor of N$_\text{TB}$. 

Biaxiality is also 
regarded as a key factor to get spontaneous chiral symmetry 
breaking from first 
principles \cite{ref35}. These symmetry arguments are supported by 
recent phenomenological analysis of modulated nematic structures 
using generalized Landau--de~Gennes--Ginzburg theory, where the 
key ingredients were couplings between 
the alignment 
tensor field and steric polarization \cite{ref22}. In this theory
the N$_\text{TB}$ phase, described by a 
locally uniaxial distortion of the director field, appears less 
stable than its locally biaxial counterpart,
\textit{i.e.} 
where full spectrum of distortions of the alignment tensor are 
taken into account. Following this direction we 
extend the GLF model  
to study the effect of molecular biaxiality of bent--core 
molecules on stability of N$_\text{TB}$ and of competing 
nematic phases, including biaxial one. For this purpose we replace 
the local twist--bend spatial modulation of the director field 
(\ref{nn}) by its biaxial counterpart, given by the local 
alignment tensor. We also let the 
molecular biaxiality of banana--shaped 
molecules to enter not only through their "V" shape but also 
through biaxiality of molecular arms. This extension  
allows to treat arm molecular biaxiality as an extra parameter 
characterizing bent--shaped molecules, in addition
to the bend angle.

This paper is organized as follows. In the second Section we 
define extension of the GLF model and underline its important 
features. Next we interpret acquired results in the third 
Section. The last 
Section is devoted to a short discussion. 
%
%
\section{The model}
\subsection{Molecular geometry and director profiles}
\begin{figure}[h!]
\centering
    \includegraphics[scale=.77]{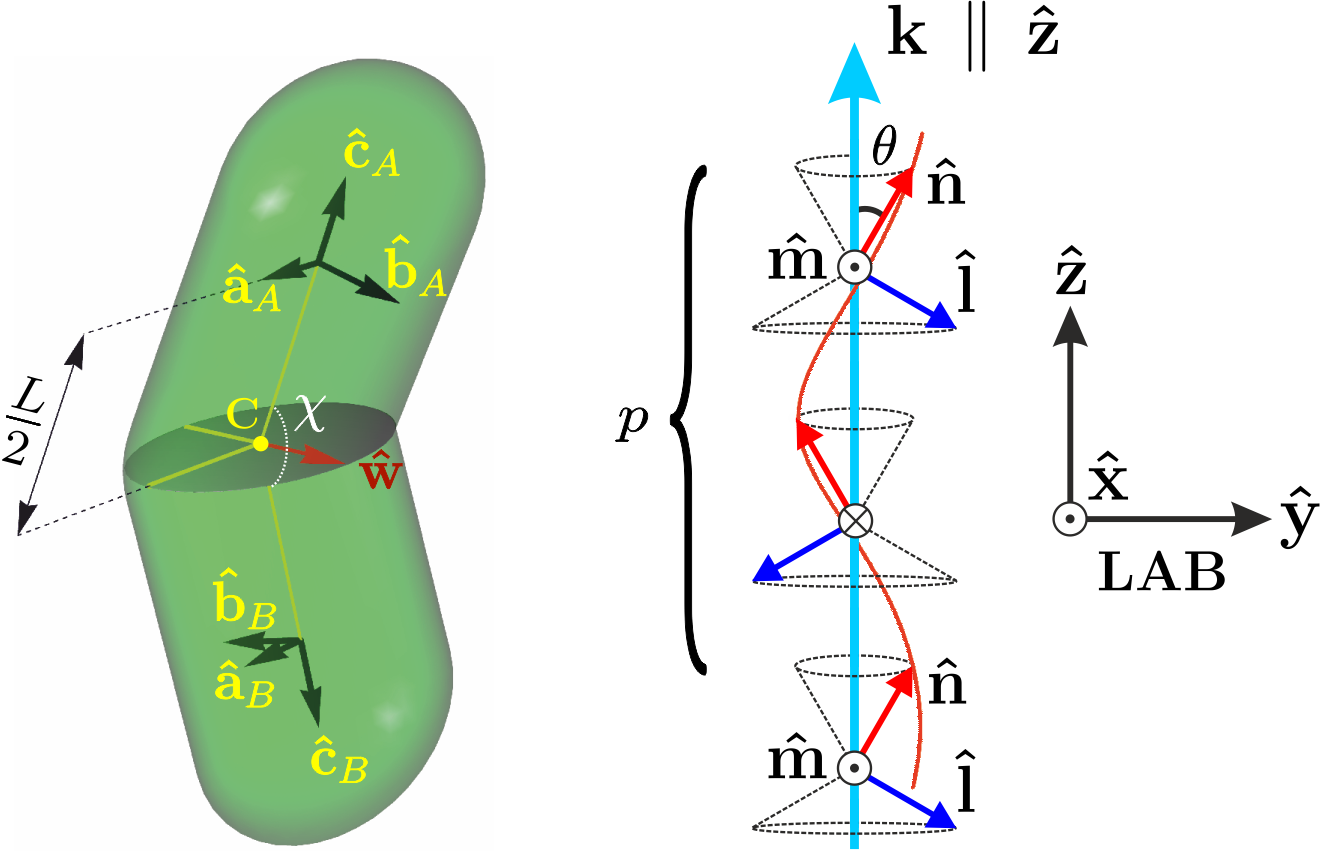}
\caption{(Color online) On the left side is a schematic 
representation of 
extended biaxial model for bent--shaped molecule with both arms 
of equal  
length $L$. Molecular basis for arm A is given by  
orthonormal tripod of 
vectors $\Omega_A=\{\mathbf{\hat{a}}_A,\mathbf{\hat{b}}_A,\mathbf
{\hat{c}}
_A\}$, and for arm B  by  $\Omega_B=\{\mathbf{\hat{a}}_B,\mathbf
{\hat{b}}
_B,\mathbf{\hat{c}}_B\}$. $C_2$ axis of  molecule is along  
unit 
vector 
$\mathbf{\hat{w}}$, attached at point C. Bend angle 
is 
denoted as $\chi$. Sketch on the right--hand side of the figure 
shows a visualization of local directors used in the mean--field \emph{ansatz} for N$_\text{TB}$ structure, where local
director basis 
$\{\mathbf{\hat{n}},\mathbf{\hat{m}},\mathbf{\hat{l}}\}$ 
on each arm precesses on a cone of pitch $p$ and tilt 
angle $\theta$ between primary director $\mathbf{\hat{n}}$ and wave
vector 
$\mathbf{k}$. }
\label{fig:scheme1}
\end{figure}

Here we keep parametrization \cite{ref31} for molecular reference frames,
where two mesogenic arms A and B, each of length $L$, are joined at the bend 
angle $\chi$ (Figure \ref{fig:scheme1}). At the midpoint of each arm a  
molecular basis is placed. The unit vector $\mathbf{\hat{w}}$ 
at the center (C) of the particle describes molecular $C_{\text{2}}$ 
axis. 
In line with the present
model of the N$_\text{TB}$ phase \cite{ref25,ref31} we 
generalize the local uniaxial \emph{ansatz}, represented by 
the main director (\ref{nn}), 
by its local biaxial counterpart (kept from the start in the 
variational \emph{ansatz} for local environment of the molecule).
That is, we assume two other 
directors to follow the precession of $\mathbf{\hat{n}}(z)$ (Figure \ref
{fig:scheme1}): 
\begin{equation}\label{mm}
\mathbf{\hat{m}}(z) =[\cos (\phi ),\sin (\phi ),0],
\end{equation}
\begin{equation}\label{ll}
\mathbf{\hat{l}}(z) \equiv \mathbf{\hat{n}}(z) \times \mathbf{\hat{m}}(z) = 
[-\cos (\theta ) \sin (\phi ),\cos (\theta ) \cos (\phi ),-\sin (\theta )].
\end{equation}
As for GLF the parametrization (\ref{nn}-\ref{ll}) 
permits the N$_{\text{TB}}$ phase for 
$0^\circ<\theta<90^\circ$ and finite pitch $p$, with wave
vector $\mathbf{{k}}$ being parallel to the $\mathbf{\hat{z}}$ axis.

\subsection{Formulation of the mean--field potential}
In the first step we extend the GLF model by introducing molecular biaxiality on both of the arms of the molecule, 
which is achieved via second--rank, $3\times3$, symmetric and traceless tensor $\mathbf{Q}$. The basis of the $\mathbf{Q}$ tensors, defined with respect to the orthonormal right--handed tripod, say $\{\mathbf{\hat{x}},\mathbf{\hat{y}},\mathbf{\hat{z}} \}$,   comprises both uniaxial $\mathbf{Q}_\text{U}$ and biaxial $\mathbf{Q}_\text{B}$ parts, given in the general form as \cite{ref46}:
\begin{equation}
\mathbf{Q}_\text{U}(\mathbf{\hat{z}})\overset{\text{def}}{=}\frac{1}{\sqrt{6}}(3\mathbf{\hat{z}}\otimes\mathbf{\hat{z}}-\mathbb{1} ),
\label{qu}
\end{equation}
\begin{equation}
\mathbf{Q}_\text{B}(\mathbf{\hat{x}},\mathbf{\hat{y}})\overset{\text{def}}{=}\frac{1}{\sqrt{2}}(\mathbf{\hat{x}}\otimes\mathbf{\hat{x}}-\mathbf{\hat{y}}\otimes\mathbf{\hat{y}}),
\label{qb}
\end{equation}
where $\otimes$ denotes the tensor product and $\mathbb{1}$ is the 
identity matrix. Taking linear combination of $\mathbf{Q}_\text{U}
$ and $\mathbf{Q}_\text{B}$ the molecular tensors for each arm 
are 
now defined as:
\begin{equation}\label{tensors}
\mathbf{Q}(\Omega_i)\overset{\text{def}}{=}\mathbf{Q_U}(\mathbf
{\hat{c}}_i)+ \lambda \sqrt{2} \mathbf{Q_B}(\mathbf{\hat{a}}
_i,\mathbf{\hat{b}}_i),
\end{equation}
where the $\lambda$ parameter is a measure of the arm's biaxiality
and where
$\Omega_i=\{\mathbf{\hat{a}}_i,\mathbf{\hat{b}}_i,\mathbf{\hat{c}}
_i\}$ is the molecular right--handed tripod 
attached  to arm $i=\text{A,B}$ (Figure \ref{fig:scheme1}). 
Please observe that the GLF model \cite{ref31} corresponds
to $\lambda=0$. In addition we should mention that the $\mathbf{Q}
(\Omega_i)$ tensor can be linked to the 
diagonal elements of molecular 
polarizability tensor \cite{ref38} for the arm $i$. 

The next step is decomposition of the tensor 
$\langle\mathbf{{Q}}(\Omega_j)\rangle\overset{\text{def}}{=}
\mathbf{\bar{Q}}(\mathbf{\hat{n}}(\text{R}_j),
\mathbf{\hat{m}(\text{R}_j)},
\mathbf{\hat{l}}(\text{R}_j))
\equiv \mathbf{\bar{Q}}(\text{R}_j)$, 
which is obtained from 
Eq.~(6) by performing 
thermodynamic average.  In the basis (\ref{qu},\ref{qb}) it 
reads:
\begin{eqnarray} \label{alignmentQ}
 \mathbf{\bar{Q}}(\text{R}_j) &=& \mathbf{\bar{Q}}_U(\text{R}_j) + \mathbf{\bar{Q}}_B(\text{R}_j)\nonumber\\ 
 &=& q_0\mathbf{Q}_\text
{U}(\mathbf{\hat{n}}(\text{R}_j)) + q_2 \mathbf{Q}_\text{B}
(\mathbf{\hat{m}}(\text{R}_j),\mathbf{\hat{l}}(\text{R}_j)),
\end{eqnarray}
where 
$\left\{\mathbf{\hat{n}}(\text{R}_j),
\mathbf{\hat{m}}(\text{R}_j),
\mathbf{\hat{l}}(\text{R}_j)
\right\}$ are the three local 
directors 
at the 
position $\text{R}_j$ of the midpoint of the $j{\text{--th}}$ 
arm 
($j=\text{A,B}$), and where $q_0$ and $q_2$ are the uniaxial 
and biaxial order parameters of an arm, 
respectively. The local directors are identified with 
eigenvectors of  
the $\mathbf{\bar{Q}}$ tensor and the corresponding 
eigenvalues \cite
{ref39} are given by $\mu_m=-q_0/\sqrt{6}+q_2/\sqrt{2}$, 
$\mu_l=-q_0/\sqrt{6}-q_2/\sqrt{2}$ and $\mu_n=-\mu_m-\mu_l=\sqrt
{2/3}q_0$. From this perspective the locally 
isotropic phase is met when all three 
eigenvalues of $\mathbf{\bar{Q}}$ are equal, which means $\mathbf
{\bar{Q}}\equiv \mathbf{0}$. For the locally $D_{\infty \text{h}
}$--symmetric uniaxial states two out of three eigenvalues of 
$\mathbf{\bar{Q}}$ are equal, \textit{i.e.}, $q_0\neq0$, $q_2=0$ 
or $q_0\neq0$, $q_2=\sqrt{3}q_0$ or $q_0\neq0$, $q_2=-\sqrt{3}q_0
$. In the general case,  $\mathbf{\bar{Q}}$ has three different 
real eigenvalues that correspond to the local, $D_{\text{2h}}
$--symmetric biaxial state.

Finally, we can write down a full mean--field potential as a 
natural extension of that proposed by GLF \cite{ref31}.
It reads
\begin{equation}
H_{\text{MF}}(\Omega)=-\varepsilon\Tr{[\mathbf{Q}(\Omega_A)
\cdot \mathbf{\bar{Q}}(R_\text{A}) +\mathbf{Q}(
\Omega_B)\cdot \mathbf{\bar{Q}}(R_\text{B})]},
\label{eq:hamiltonian}
\end{equation}
where $\varepsilon$ is the coupling constant, '$\cdot$' denotes 
matrix multiplication and $\Omega$ stands for the molecular orientation 
expressed in terms of  Euler angles that define this orientation in a local $\{\mathbf{\hat{n}},\mathbf{\hat{m}},\mathbf{\hat{l}}\}$ frame. The corresponding mean--field 
equilibrium free energy per particle resulting from orientational degrees of freedom is given by:
\begin{equation}\label{free}
f=q^2_0+q^2_2-t^* \ln{Z},
\end{equation}
where  $Z$  is the orientational one--particle partition function:
\begin{equation}
Z=\int e^{-H_{\text{MF}}(\Omega)/t^*} d\Omega, 
\end{equation}
and where  $t^*\overset{\text{def}}{=}k_{\text{B}}T/
\varepsilon$ 
is the (dimensionless) reduced temperature. Orientational 
averages of any 
one--particle quantity $X(\Omega)$ are calculated in a standard 
way as:
\begin{equation}
\langle X(\Omega) \rangle=\frac{1}{Z}\int X(\Omega)e^{-H_{\text{MF}}(\Omega)/t^*} d\Omega. 
\end{equation}

Taking parametric form (\ref{nn}--\ref{ll},
\ref{alignmentQ}) 
of the alignment tensor $\mathbf{\bar
{Q}}$ the equilibrium structure can be obtained by 
minimization of the free energy, Eq.~(\ref{free}), with 
respect to the order parameters  $q_0$ and $q_2$ and the
"local environment" parameters $\theta$ and $p$.
The former ones are given self--consistently by: 
\begin{equation}
q_n = \frac{1}{2} \langle q_n\left(\{\text
{R}_A\},  \Omega_A \right)  
+ q_n\left( \{\text{R}_B\}, 
\Omega_B \right) \rangle, \,\,\,\,\, n=0,2, 
\end{equation}
where the orientational averaging applies to the symmetry 
adapted functions which are given in a typical form \cite
{ref46,ref39}: 
\begin{eqnarray}
q_0( \{\text{R}_k\}, \Omega_k) &=& -\frac{1}{2} + 
\frac{3}{2}(\mathbf{\hat{n}}(\text{R}_k) \cdot \mathbf
{\hat{c}}_k)^2 \nonumber  \\ 
 && \hspace{-1cm} +\, \lambda  \sqrt{\frac{3}{2}} \left[ ( 
 \mathbf{\hat{n}
}(\text{R}_k) \cdot \mathbf{\hat{a}}_k )^2 - ( \mathbf{\hat
{n}}(\text{R}_k) \cdot \mathbf{\hat{b}}_k )^2 \right], \\
q_2(\{\text{R}_k\}, \Omega_k) &=& \frac{\sqrt{3}}{2} [ ( \mathbf
{\hat{l}} (\text{R}_k)  \cdot \mathbf{\hat{c}}_k )^2 - ( 
\mathbf{\hat{m}}(\text{R}_k) \cdot \mathbf{\hat{c}}_k )^2 ] 
\\ \nonumber 
&& \hspace{-2cm} + \lambda \sqrt{2 }\left[ ( \mathbf{\hat
{l}} (\text{R}
_k) \cdot\mathbf{\hat{a}}_k)^2 + ( \mathbf{\hat{m}} (\text
{R}_k) \cdot \mathbf{\hat{b}}_k )^2 - \frac{1}{2}( \mathbf
{\hat{n}} (\text{R}_k) \cdot \mathbf{\hat{c}}_k )^2 - \frac
{1}{2} \right],
\end{eqnarray}
and where symbol $\{\text{R}_k\} \overset{\text{def}}{=} \left\{\mathbf{\hat{n}}(\text{R}_k),
\mathbf{\hat{m}}(\text{R}_k),
\mathbf{\hat{l}}(\text{R}_k)
\right\}$
stands for the right--handed tripod of directors ($k=A,B$).

Before going further it seems appropriate 
to 
discuss similarities and differences between the present
model (\ref
{eq:hamiltonian}) and that of GLF. To this end 
we rewrite the GLF Hamiltonian in our notation:
\begin{equation}\label{HGLF}
H_{\text{MF}}^{\text{GLF}}=-\varepsilon \Tr{[\mathbf{Q_U}
(\mathbf{\hat{c}}_A)\cdot \bar{\mathbf{Q}}_\text{U}
(\text{R}_A)+ \mathbf{Q_U}(\mathbf{\hat{c}}_B)\cdot
\bar{\mathbf{Q}}_\text{U}(\text{R}
_B)]}.
\end{equation} 

The form of $\bar{\mathbf{Q}}_\text{U}$ 
and 
$\bar{\mathbf{Q}}$ tensors accounts in both models for the 
\textit{global} $D_\infty$ symmetry point group of the 
N$_\text{TB}$ phase with the (optical) 
$C_\infty$  axis parallel to the helix axis 
\cite{ref10,PhysRevLett.111.067801}.
N$_\text{TB}$ is also invariant for the 
twofold  rotation around a local vector $\mathbf{\hat{m}}$,
where $\mathbf{\hat{m}}$ is  
perpendicular to the plane 
containing the helix axis $\mathbf{\hat{z}}$ and the local 
director. This local 
$C_2$ symmetry causes that N$_\text{TB}$ is locally polar.
As $\mathbf{\hat{n}}$, $\mathbf{\hat{m}}$ and $\mathbf{k}$ 
are linearly independent the structure is also locally biaxial.

The difference between the two models lies
in primary order parameters entering 
heliconical variational \textit{ansatz} (\ref{alignmentQ}) on the N$_\text{TB}$ structure. While in the GLF model the conical twist-bend helix 
$\bar{\mathbf{Q}}_\text{U}$ is approximated by a locally uniaxial
distortion of the director field weighted by $q_0$,  our $\bar
{\mathbf{Q}}$ tensor comprises full set of the directors
(\ref{nn}--\ref{ll}) which, along with the variational parameters 
$q_0$  and  $q_2$, permits the helix to be locally biaxial. 
Both order parameters, $q_0$ and $q_2$, can be determined experimentally along 
the effective optical axis $\mathbf{k} \parallel \langle \mathbf{\hat{n}} \rangle_p$\cite{ref15,ref16,ref18}, which is an eigenvector of  $\langle\bar{\mathbf{Q}}\rangle_p$, where $\langle \dots \rangle_p$ denotes average over one period $p$ along $\mathbf{k}$. Indeed, the averaged alignment tensor $\langle\bar{\mathbf{Q}}\rangle_p$  
is diagonal, uniaxial and of zero trace, and the eigenvector $\langle\mathbf{\hat{n}}\rangle_p$ corresponds to the non--degenerate eigenvalue $\Lambda_k$, 
given by a linear combination of $q_0$ and  $q_2$:
\begin{equation}\label{lambdak}
\Lambda_k = \frac{{q_0} (3 \cos (2 \theta
   )+1)}{2
   \sqrt{6}}+\frac{{q_2} \sin
   ^2(\theta )}{\sqrt{2}}.
\end{equation}

Our extension is also important as it obeys two nematic phases, uniaxial and biaxial, both recovered for pitch $p\to \infty$, while in the GLF model only a uniaxial nematic phase is present. 
Further difference between the models concerns the treatment of molecular biaxiality, which we discuss below.
\subsection{Effective molecular shape in the nematic limit}

V--shaped molecules of both models are biaxial due to their  $C_{2v}$ symmetry.
In the GLF model they are represented  by two uniaxial arms with the bend angle $\chi$, while our model permits molecular arms to be biaxial, with arm's biaxiality controlled by $\lambda$. Clearly, in both cases the total molecule is biaxial, but the model (\ref{eq:hamiltonian}) allows for full control of a composite molecular biaxiality. In order to illustrate this, we study the effective molecular shape of the two models as seen in the nematic limit ($\phi = 0$ in equations (\ref{nn}--\ref{ll}). Since in this limit directors become positionally independent, it implies that
 $\mathbf{\bar{Q}}(R_\text{A}) = \mathbf{\bar{Q}}(R_\text{B}) \overset{\text{def}}{=} \mathbf{\bar{Q}}$ and  
\begin{equation} \label{hmfN} 
H_{\text{MF,N}}(\Omega)=-\varepsilon\Tr[ ( \mathbf{Q}(\Omega_A) + \mathbf{Q}(\Omega_B)) \cdot \mathbf{\bar{Q}} ].
\end{equation}

That is, with the nematic \emph{ansatz} the segmental mean--field model  (\ref{eq:hamiltonian}) can be mapped into single site, mean--field version of the dispersion model \cite{ref46}, where
the bent--shaped molecule is represented by an effective molecular 
quadrupolar tensor
\begin{equation}
\mathbf{Q}_{\text{mol}} \overset{\text{def}}{=} \mathbf{Q}(\Omega_A) +\mathbf{Q}(\Omega_B). 
\end{equation}

\begin{figure}[h!]
\centering
    \includegraphics[scale=0.951]{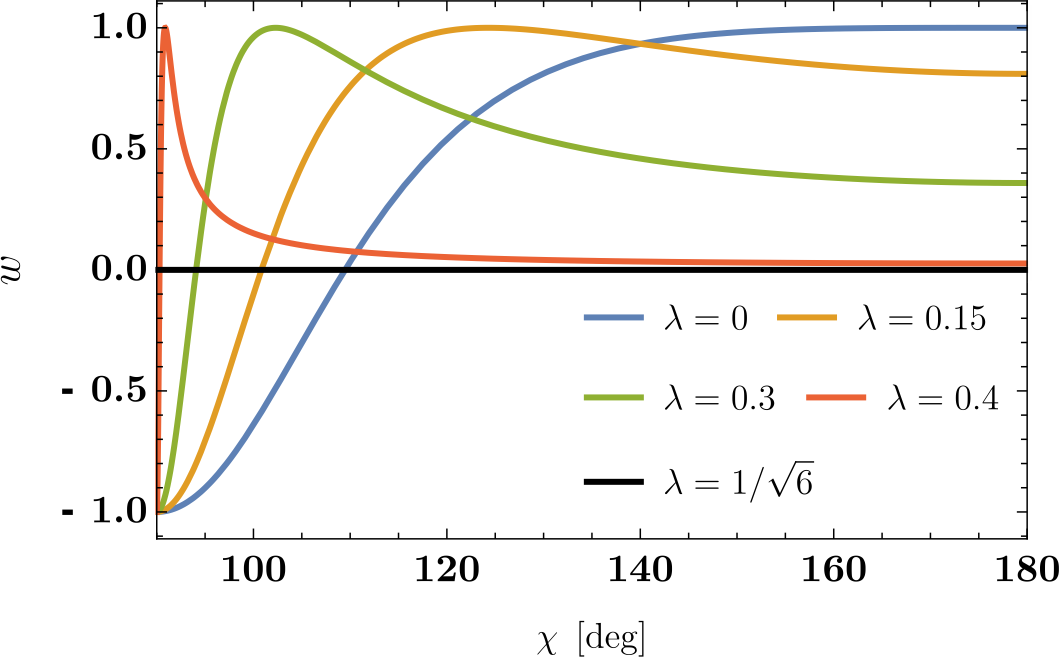}
\caption{(Color online) Molecular biaxiality parameter (\ref{wm}) in nematic phase  as a function of bend angle for $\lambda$ equal to: $0$ (blue), $0.15$ (orange), $0.3$ (green), $0.4$ (red) and $1/\sqrt{6}$ (black).}
\label{fig:wm}
\end{figure}

The $\mathbf{Q}_{\text{mol}}$ tensor is biaxial, also for the GLF model of $\lambda=0$. The biaxiality of $\mathbf{Q}_{\text{mol}}$ can be quantified by calculating the invariant, normalized biaxiality parameter $w=\sqrt[]{6} \Tr[\mathbf{Q}_{\text{mol}}^3]/
\Tr[\mathbf{Q}_{\text{mol}}^2]^{\frac{3}{2}}\,\, (w^2\le 1)$ \cite{ref46,ref39}.
It reads 
\begin{eqnarray}\label{wm}
w = \frac{3 \sqrt{6} \left(2 \lambda ^2+1\right) \lambda  \sin ^2(\chi) +\left(9-30 \lambda ^2\right) \cos ^2(\chi) -18 \lambda ^2-1}{\left(\lambda ^2 \cos (2 \chi )+7 \lambda ^2-2 \sqrt{6} \lambda  \sin ^2(\chi) +3 \cos ^2(\chi) +1\right)^{3/2}}.
\end{eqnarray}
For the uniaxial states $w^2=1$, whereas nonzero biaxiality is monitored by $w^2<1$ approaching maximal value for $w=0$. The case $w>0$ ($w<0$) corresponds to prolate (oblate) states of $\mathbf{Q}_{\text{mol}}$. The variation of $w$ with the angle between the two arms calculated from Eq.~(\ref{wm}) for a  selection of the  values of the  $\lambda$ parameter is given in Figure \ref{fig:wm}.

For the GLF model ($\lambda=0$) the V--shaped molecule exhibits effectively disc--like uniaxial shape at $\chi=90^\circ$, which evolves to rod--like uniaxial one at $\chi=180^\circ$. The curve in $(w,\chi)$ plane passes through zero, the point of maximal molecular biaxiality, when the bend angle is equal to the tetrahedral
value ($\chi = \arccos(-1/3) \cong109.47^\circ$). In spite of this molecular biaxiality the GLF \emph{ansatz} (\ref{HGLF}) permits only uniaxial structures, which excludes \textit{e.g.} the biaxial nematic phase.

For $\lambda>0$ the effective molecular biaxiality can be made less
dependent on $\chi$ and already for $\lambda \gtrsim 0.15$ the arm--induced biaxiality starts prevailing over.  In the limit of maximally biaxial arms 
($\lambda=1/\sqrt6$), the effective molecule becomes maximally biaxial irrespective of the
angle between the arms. Hence, the simple mean--field model (\ref{eq:hamiltonian}) with only two molecular parameters,  $\lambda$  and $\chi$, allows to control almost independently molecular  anisotropy and the angle between the arms. They both seem primary to liquid crystal behavior of bent--core, dimeric and trimeric mesogens, especially in view that compounds composed of these molecules are also candidates to exhibit the elusive biaxial nematic phase.

\section{Results}
 In what goes after, we use the following notation for phases: N$_{\text{U}}$ for the uniaxial nematic, N$_{\text{B}}$ for the biaxial nematic, N$_\text{TB}$ for the twist--bend nematic with heliconical uniaxial \emph{ansatz} ($\mathbf{\bar{Q}}_U$), N$_{\text{TB,B}}$ for the twist--bend nematic with heliconical biaxial \emph{ansatz} ($\mathbf{\bar{Q}}$) and $\text{Iso} $ for the isotropic phase.
 
\begin{figure}[h!]
\centering
\subfigure[]{%
\includegraphics[width=.45\textwidth]{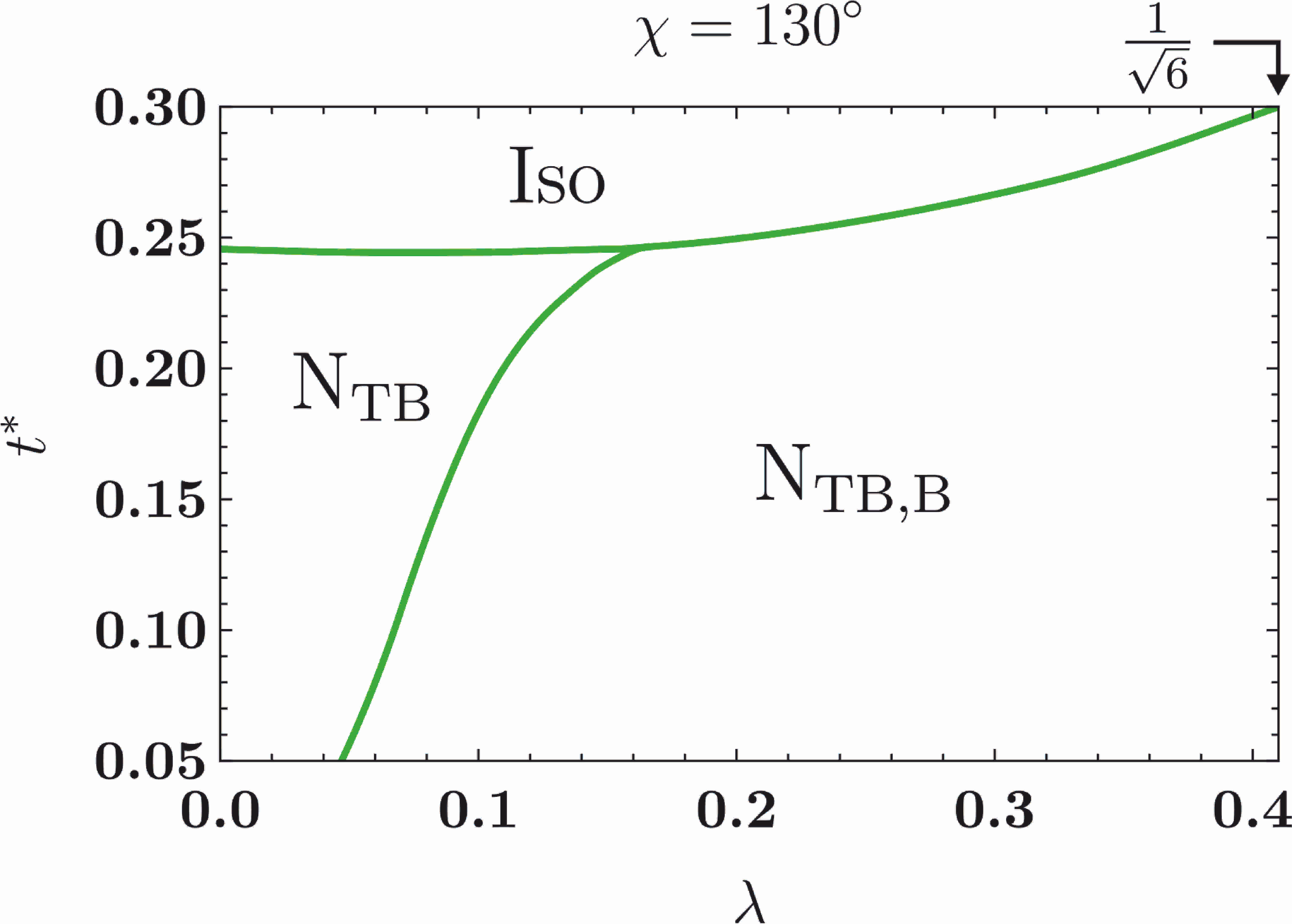}}\\
\subfigure[]{%
\includegraphics[width=.45\textwidth]{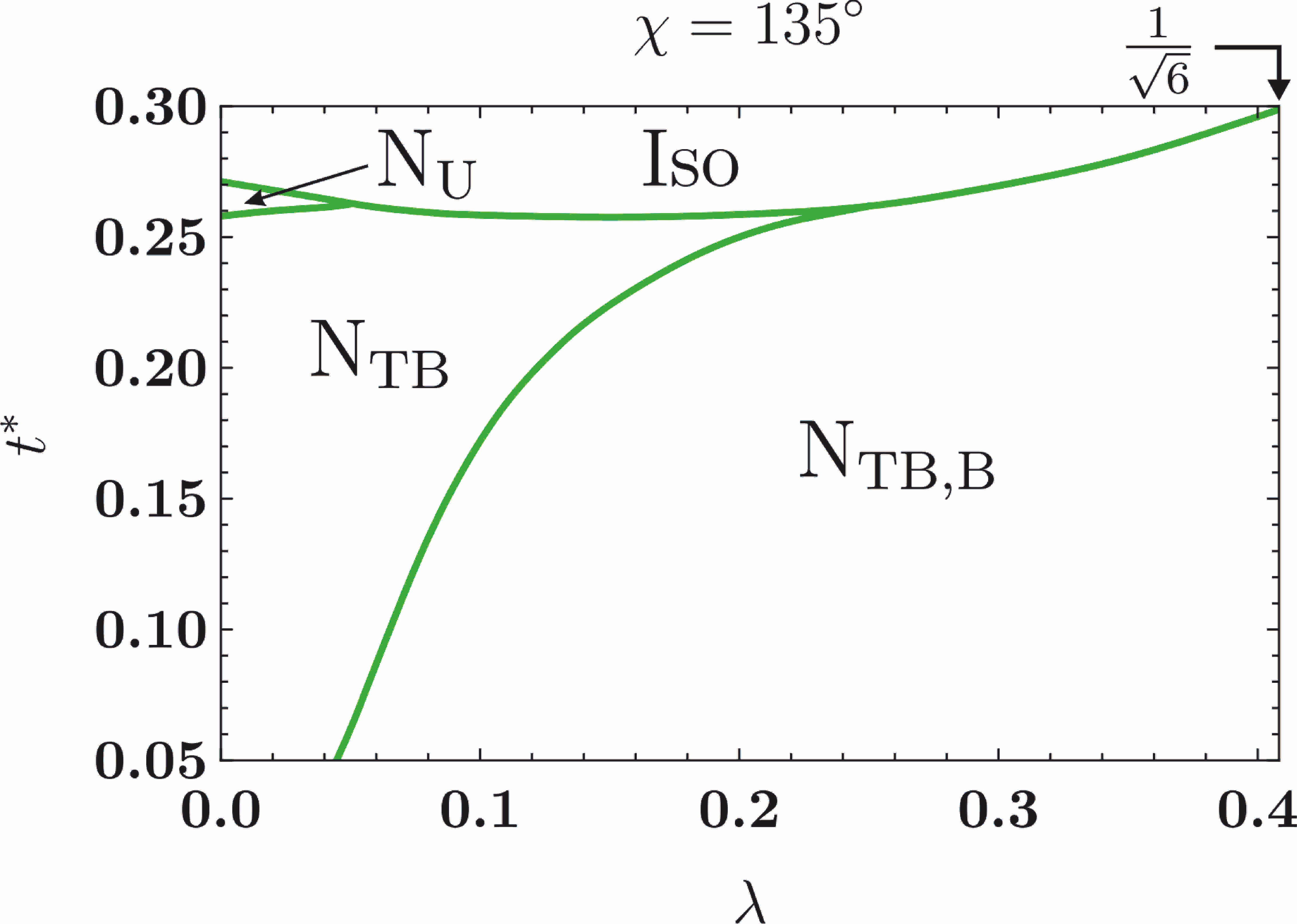}}\\
\subfigure[]{%
\includegraphics[width=.45\textwidth]{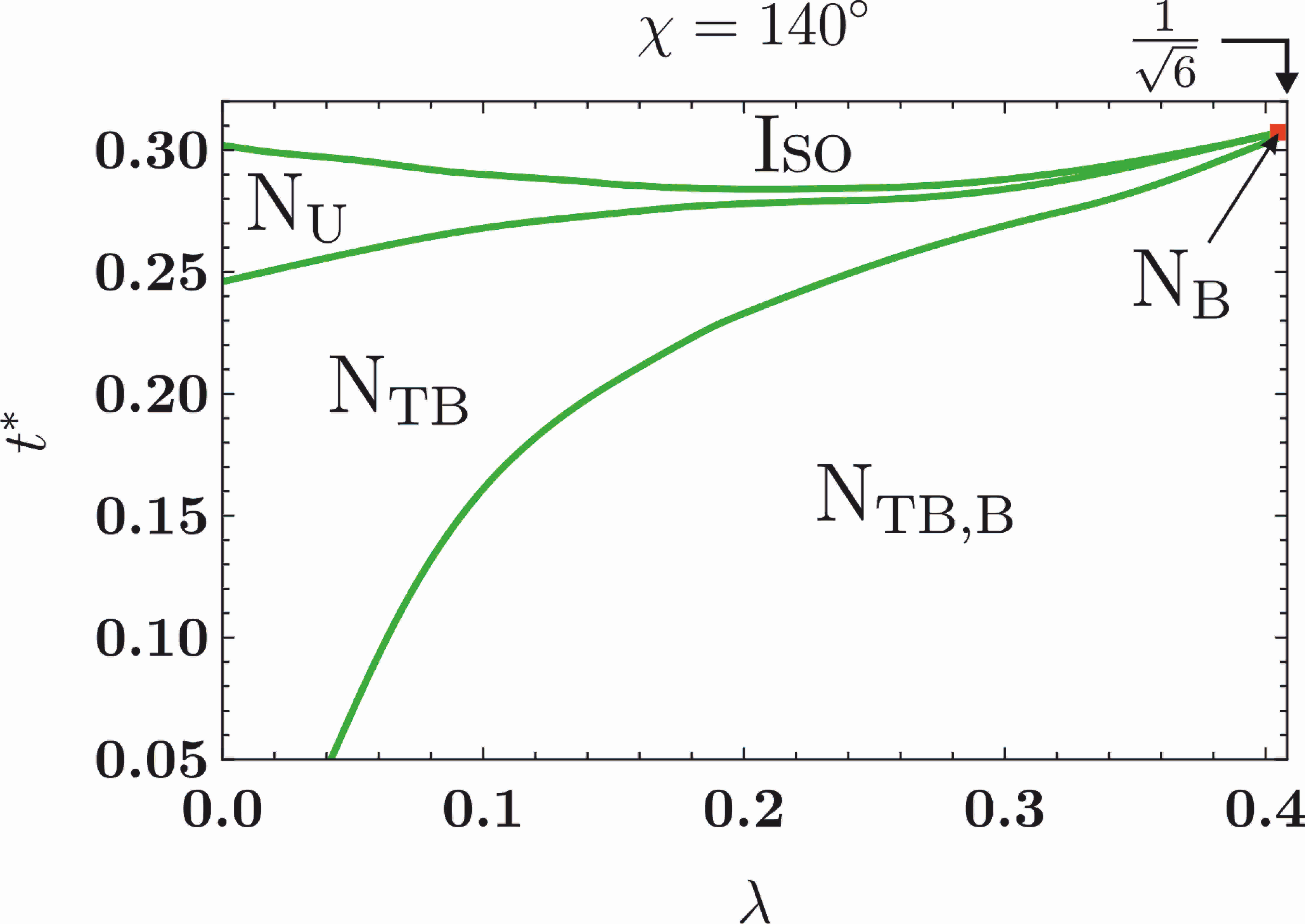}}
\label{fig:phasediagram}
\caption{(Color online) Phase diagrams on $(\lambda, t^*)$ plane for bend angle equal to (a) $130^\circ$, (b) $135^\circ$ and (c) $140^\circ$. Red square in panel (c) represents tiny region where stable $\text{N}_{\text{B}}$ phase occurs; its area  enhances with further increase of the bending angle $\chi$. }\label{fig:phase_diagram}
\end{figure}

Phase diagrams for banana--shaped biaxial molecules for three typical values:  $\text{130}^{\circ} ,\text{135}^{\circ}$ and $\text{140}^{\circ}$ of the bend angle $\chi$ are presented in Figure \ref{fig:phase_diagram}. For $\chi=130^\circ$ three phases:  N$_\text{TB}$, N$_{\text{TB,B}}$, and $\text{Iso} $ become stable  for $\lambda$ ranging from 0 to the so--called self--dual point \cite{ref46}, where  $\lambda=1/\sqrt{6}$. In Figure \ref{fig:phase_diagram}a the  N$_{\text{TB,B}}$ phase becomes stable below the uniaxial $\text{N}_\text{TB}$  with  phase sequence: $\text{Iso} \rightarrow \text{N}_\text{TB} \rightarrow \text{N}_{\text{TB},\text{B}}$, or directly  below $\text{Iso}$  through sequence: $\text{Iso} \rightarrow \text{N}_{\text{TB},\text{B}}$. Widening the bend angle (see Figures \ref{fig:phase_diagram}b and \ref{fig:phase_diagram}c) leads to stable regions of  N$_{\text{U}}$ and even N$_{\text{B}}$ phases, although N$_{\text{B}}$ is visible only in a very narrow region of $\lambda$ (in the closest vicinity of the self--dual point) and for $\chi=140^\circ$. Apart from phase transitions that are present for $\chi=130^\circ$ we can identify the following sequences: $\text{Iso} \rightarrow \text{N}_{\text{U}} \rightarrow \text{N}_\text{TB} \rightarrow \text{N}_{\text{TB},\text{B}}$, $\text{Iso} \rightarrow \text{N}_{\text{U}} \rightarrow \text{N}_{\text{B}} \rightarrow \text{N}_{\text{TB},\text{B}}$ and $\text{Iso} \rightarrow \text{N}_{\text{B}} \rightarrow \text{N}_{\text{TB},\text{B}}$. Broadening of regions of the nematic phases, both uniaxial and biaxial, starts with further widening of the bend angle ($\chi>140^\circ$), where the more ordered twist--bend nematics are moved to lower temperatures, similar to the tetrahedratic nematic phases \cite{ref40,ref41,ref42}. Interestingly, for $\chi\lesssim 140^\circ$ the nematic twist--bend nematic phase is always more stable than $N_B$.

\begin{figure}[ht!]
\centering
\includegraphics[scale=.176]{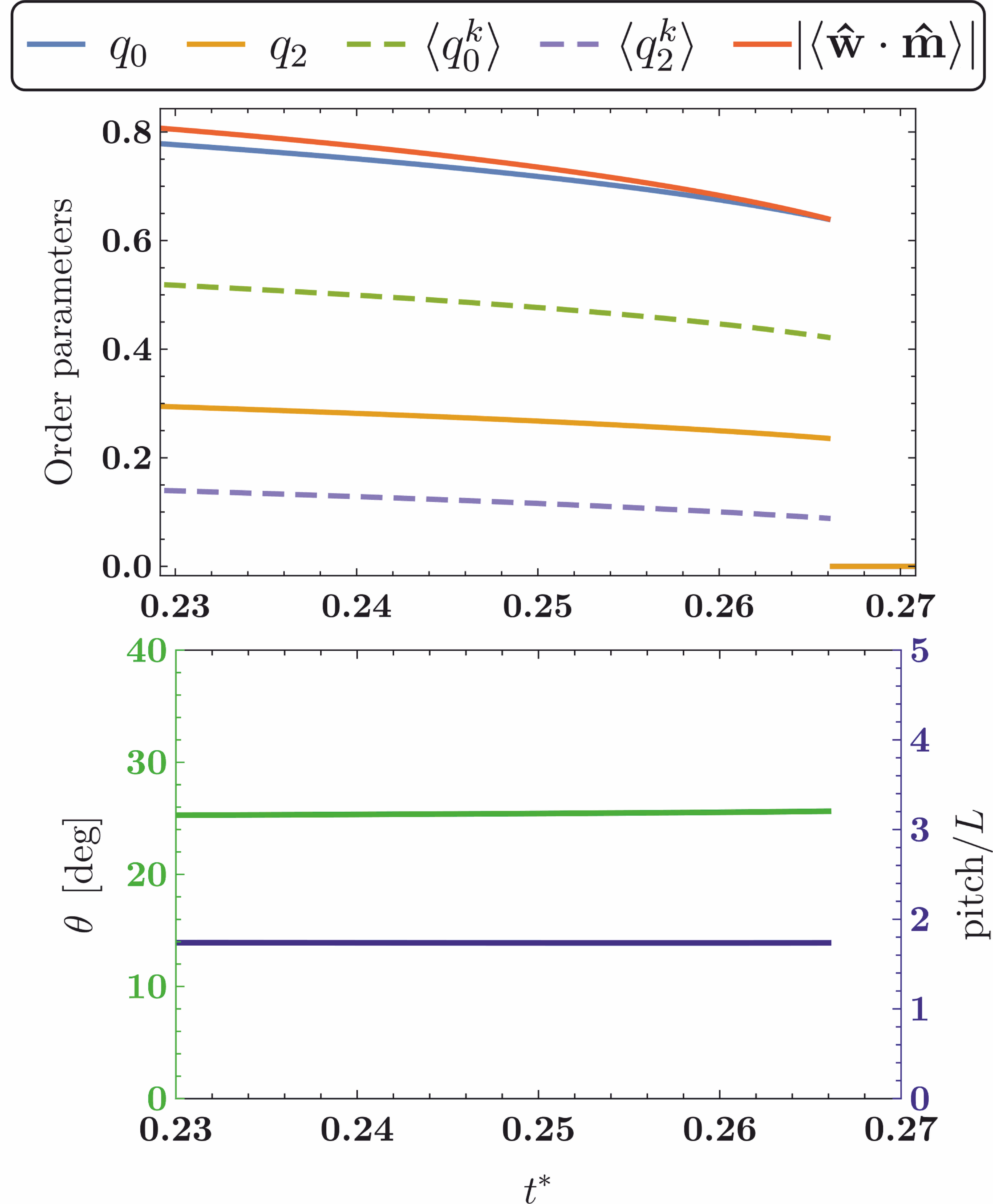}
\caption{(Color online) Behavior of order parameters, tilt angle and pitch for phase transition $\text{Iso} \rightarrow \text{N}_{\text{TB},\text{B}}$ when  $\lambda=0.3$ and $\chi = 130^{\circ}$. $q_0$ and $q_2$ are obtained by direct minimization of the  free energy (\ref{free}), $\langle q_0^k \rangle$ and $\langle q_2^k \rangle$ are uniaxial and biaxial order parameters calculated with respect to the wave vector $\mathbf{k}$, and $|\langle\mathbf{\hat{w}} \cdot \mathbf{\hat{m}} \rangle|\overset{\text{def}}{=}|\langle\mathbf{\hat{w}} \cdot \mathbf{\hat{m}}(z=\text{R}_C) \rangle|$ is the  modulus of polar order parameter. }
\label{fig:phase1}
\end{figure}

In order to better understand the identified phases, we have analyzed temperature variations of uniaxial ($q_0$) and biaxial ($q_2$) order parameters, tilt angle ($\theta$) and pitch ($p$) for the selected cases. Additionally, we have calculated the  order parameter  $\langle \mathbf{\hat{w}} \cdot~\mathbf{\hat{m}}(z=\text{R}_C) \rangle$,  which gives signature of polar order in the system, and hence allows to identify a nematic twist--bend phases. Figure \ref{fig:phase1} shows exemplary results for  the $\text{Iso} \rightarrow \text{N}_{\text{TB},\text{B}}$ phase sequence, where discontinuity in all parameters indicates on the first order phase transition between these phases. The tilt angle in $\text{N}_{\text{TB}}$ tends to $\theta =25^\circ$ and pitch is almost constant, smaller than the length of a stretched molecule.

\begin{figure}[h!]
\centering
\includegraphics[scale=.176]{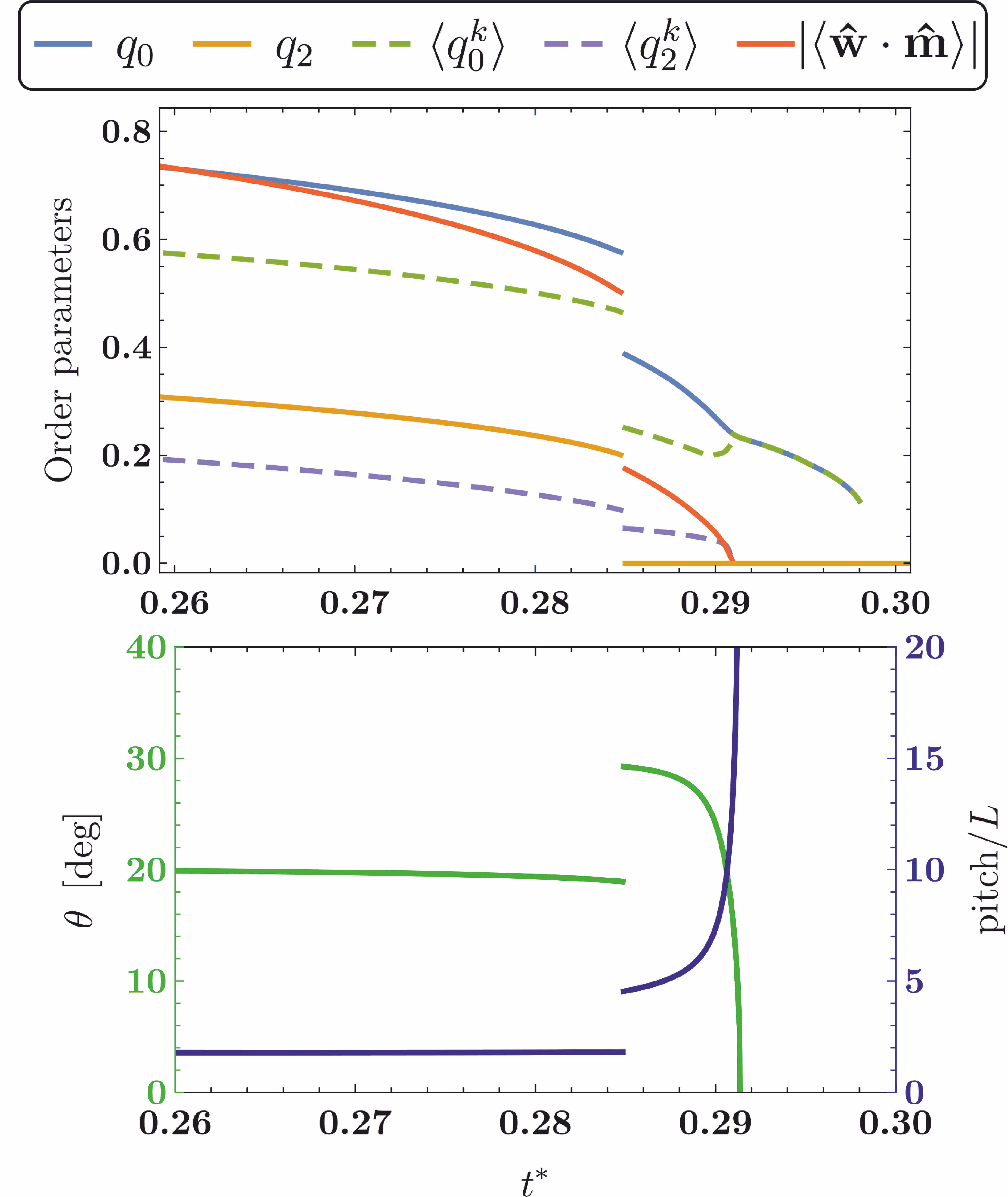}
\caption{(Color online) Behavior of order parameters, tilt angle and pitch for  $\text{Iso} \rightarrow \text{N}_{\text{U}} \rightarrow \text{N}_{\text{TB}} \rightarrow \text{N}_{\text{TB},\text{B}}$  phase transitions when $\lambda=0.36$ and $\chi = 140^{\circ}$. For further  details  see caption to Figure \ref{fig:phase1}. }
\label{fig:phase2}
\end{figure}

These results are very close to the exact value for $\theta$ that can be obtained for the ground state $(t^*=0)$: 
\begin{equation}
\theta(t^*=0)=\frac{1}{2}(180^\circ - \chi).  \label{eq:theta_limit}
\end{equation}
Indeed, substitution of $\chi=130^\circ$ gives 25$^\circ$ for $\theta$ in this limit. The relation (\ref{eq:theta_limit}), being valid for arbitrary bend angle $\chi$, is also regained for $\theta$  in the bottom panel of Figure \ref{fig:phase2}. As concerning pitch of the twist--bend phase it should never exceed $4 L$ in the ground state, but can be larger than this value for high temperatures. This is illustrated in Figure \ref{fig:phase2}, where pitch becomes divergent in the vicinity of $\text{N}_{\text{TB}} \leftrightarrow \text{N}_{\text{U}}$ phase transition.
\begin{figure}[h!]
\centering
\includegraphics[scale=.176]{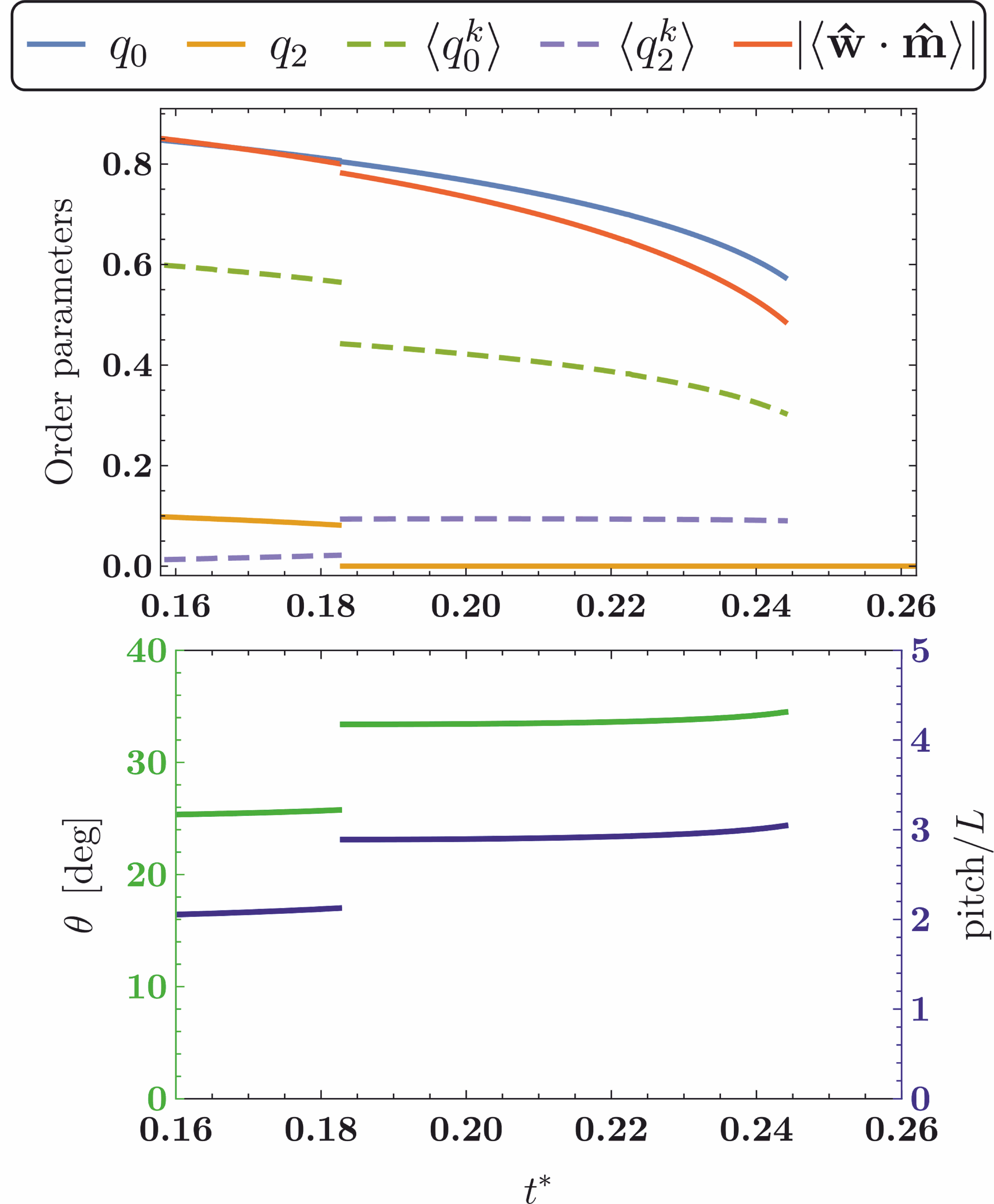}
\caption{(Color online) Behavior of order parameters, tilt angle and pitch for  phase transitions  $\text{Iso} \rightarrow \text{N}_{\text{TB}} \rightarrow \text{N}_{\text{TB},\text{B}}$ when $\lambda=0.1$ and $\chi = 130^{\circ}$. For further  details see caption to Figure \ref{fig:phase1}.}
\label{fig:phase3}
\end{figure}
\begin{figure}[h!]
\centering
\includegraphics[scale=.176]{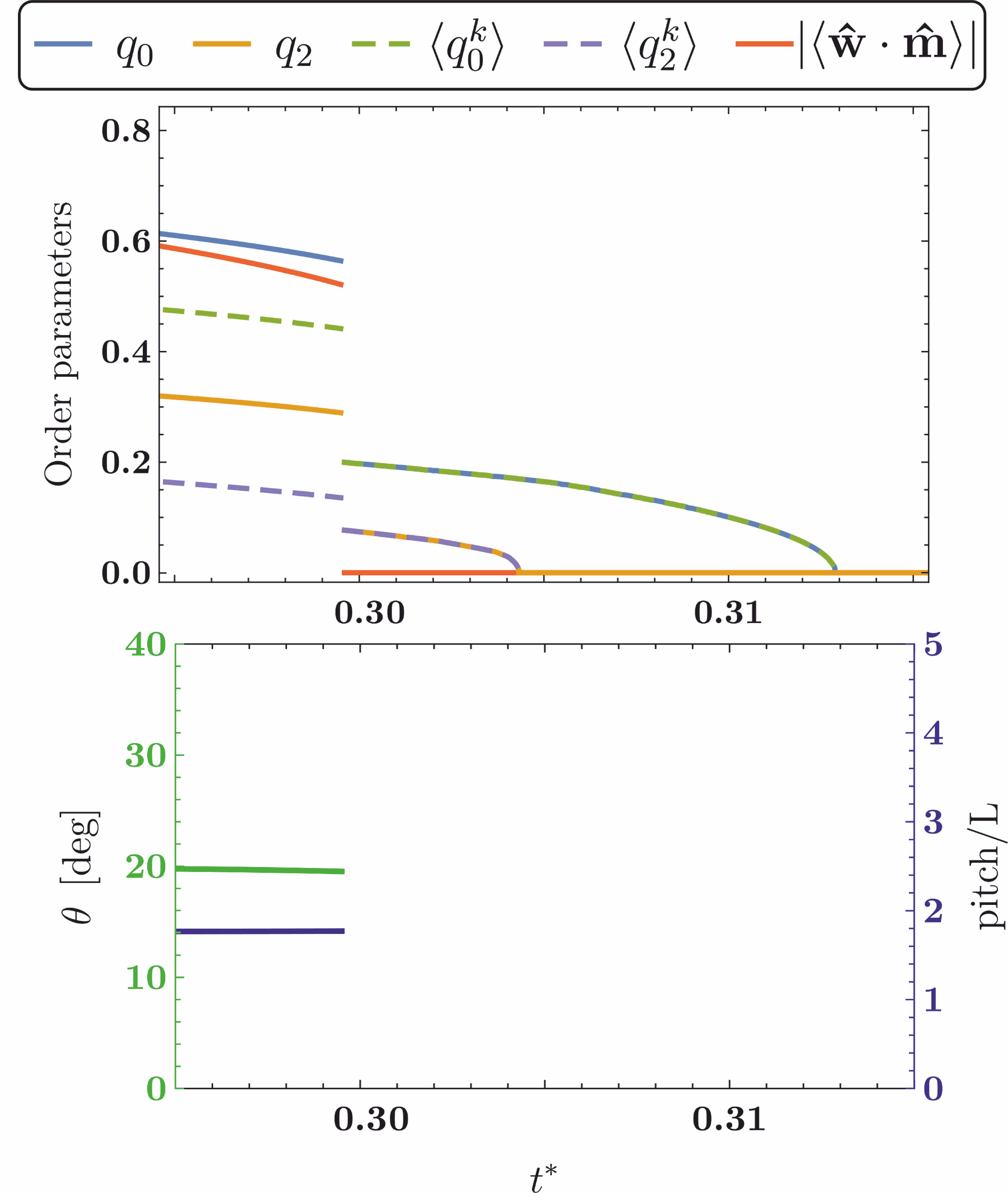}
\caption{(Color online) Behavior of order parameters, tilt angle and pitch for the sequence of phase transitions $\text{Iso} \rightarrow \text{N}_{\text{U}} \rightarrow \text{N}_{\text{B}} \rightarrow \text{N}_{\text{TB},\text{B}}$ when $\lambda=0.408$ and $\chi = 140^{\circ}$. For further details see caption to Figure \ref{fig:phase1}.}
\label{fig:phase4}
\end{figure}
\begin{figure}[h!]
\centering
\includegraphics[scale=.176]{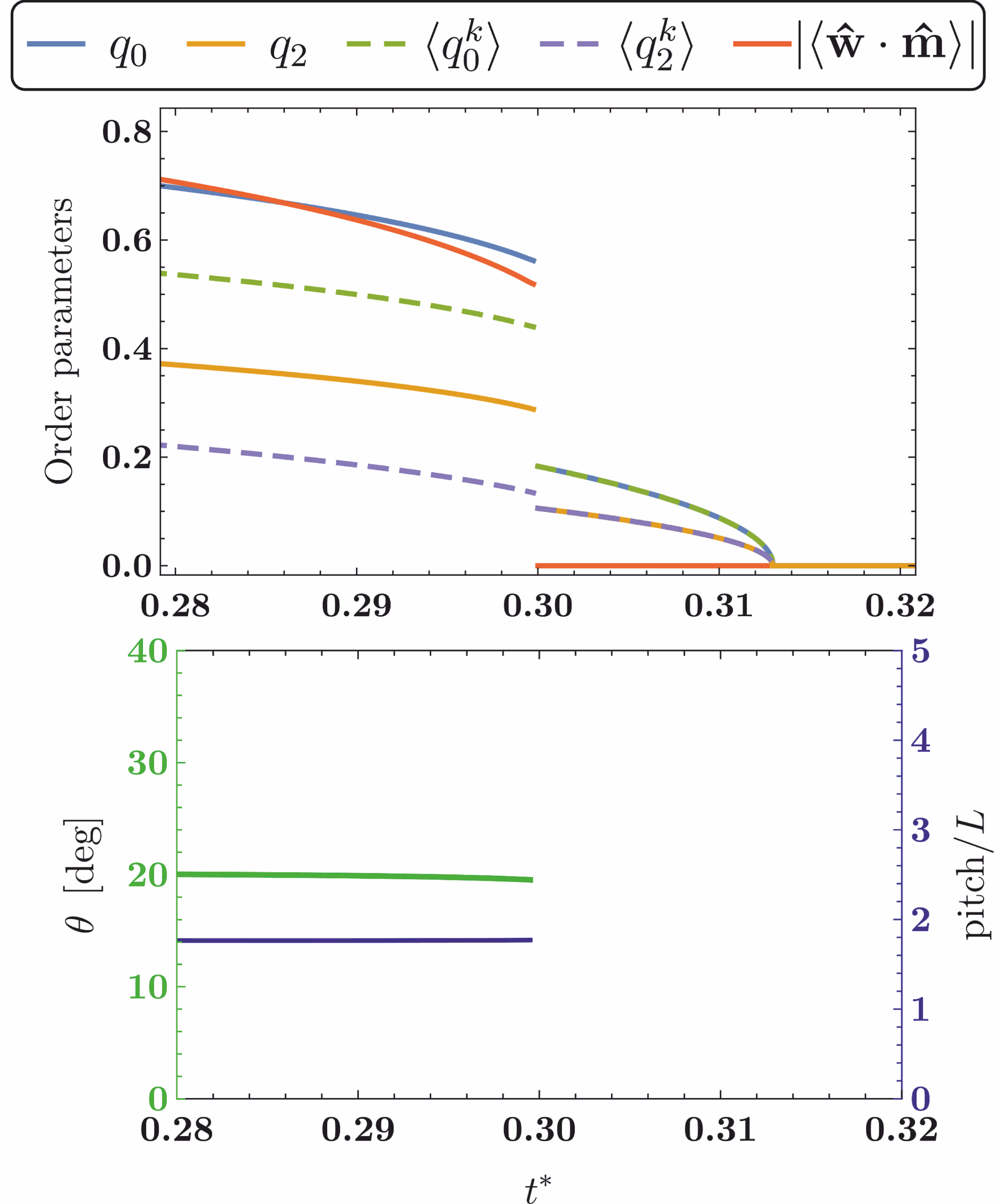}
\caption{(Color online) Behavior of order parameters, tilt angle and pitch for the successive phase transitions: $\text{Iso} \rightarrow \text{N}_{\text{B}} \rightarrow \text{N}_{\text{TB},\text{B}}$ when $\lambda=1/\sqrt{6}$ and $\chi = 140^{\circ}$. For further  details  see caption to Figure \ref{fig:phase1}.}
\label{fig:phase5}
\end{figure}

The phase diagram (Figure \ref{fig:phase_diagram}c) is especially rich in sequence of phase transitions. More specifically, we can identify first order phase transitions between $\text{N}_{\text{TB,B}}$ and $\text{N}_{\text{TB}}$ and between $\text{N}_{\text{U}}$ and Iso phases with discontinuity in order parameters (Figure \ref{fig:phase2}), and second order phase transition between $\text{N}_{\text{TB}}$ and $\text{N}_{\text{U}}$.  The second--order nature of the $\text{N}_{\text{TB}}\leftrightarrow\text{N}_{\text{U}}$ transition can be recognized from  temperature variations of conical angle and pitch, where the first goes continuously to zero while the latter diverges at the transition point. 
We also calculate mean values of uniaxial ($\langle q_0^k \rangle$) and biaxial ($\langle q_2^k \rangle$) order parameters 
with respect to the optical axis ($\mathbf{\hat{k}}=\mathbf{k}/k$) reference 
frame: $\{ \text{R}_{\mathbf{\hat{k}}} \} \overset{\text{def}}{=} \{ \mathbf{\hat{k}}, \mathbf{\hat{m}}(z=\text{R}_C),\mathbf{\hat{k}}\times \mathbf{\hat{m}}(z=\text{R}_C) \}$.
The following averages need to be determined:
\begin{equation}
\langle q_n^k \rangle = \frac{1}{2} \langle q_n\left( \{ \text{R}_{\mathbf{\hat{k}}} \}, \Omega_A \right) + q_n\left( \{ \text{R}_{\mathbf{\hat{k}}} \}, \Omega_B \right) \rangle, \,\,\,\,\, n=0,2 . 
\end{equation}
In the homogeneous $\text{N}_{\text{U}}$ phase we expect $\langle q_0^k \rangle = q_0$, which should hold for any non--tilted phase \cite{ref36,ref43,ref44}. The same relation is fulfilled by the mean value of biaxial order parameters $q_2$ and $\langle q_2^k \rangle$ in the $\text{N}_{\text{B}}$ phase. Note, however, discrepancies between the order parameters of twist--bend phases calculated in various reference frames. Locally in the arm reference frames $q_2$ is zero in the $\text{N}_{\text{TB}}$ phase, while in the $\mathbf{\hat{k}}$--frame both $\langle q_0^k \rangle$ and $\langle q_2^k \rangle$ are nonzero for any twist--bend phase. 

Further examples of  phase sequences are presented in Figures \ref{fig:phase3}--\ref{fig:phase5}, where $\text{N}_{\text{B}} \leftrightarrow \text{N}_{\text{TB},\text{B}}$ phase transition is of first order, while $\text{N}_{\text{U}} \leftrightarrow \text{N}_{\text{B}}$ and $\text{Iso} \leftrightarrow \text{N}_{\text{B}}$ are of second order \cite{ref36,ref37,ref38,ref39,ref40,ref41,ref42,ref45,ref46}. Interestingly, the conical angle and the pitch in  $\text{N}_{\text{TB}}$ are \emph{ca.} 1.3 to 1.5 and 0.8 to 2.5 times larger, respectively, than those in $\text{N}_{\text{TB,B}}$ (see Figure \ref{fig:phase2} and Figure \ref{fig:phase3}). Both twist--bend nematic phases are strongly polar, however favorably more polar is the $\text{N}_{\text{TB,B}}$ phase. Additionally, it is also noticeable that depending on the value of bending angle $\chi$ and biaxiality parameter $\lambda$ the transition between $\text{N}_{\text{TB,B}} \leftrightarrow \text{N}_{\text{TB}}$ may be characterized by significant (Figure \ref{fig:phase2}) or slight (Figure \ref{fig:phase3}) decrease in values of polar order parameter. 

Figure \ref{fig:phase4} shows a magnification of the phase sequence for $\chi=140^\circ$ in the proximity of the self--dual point (red square in Figure \ref{fig:phase_diagram}c). A noticeable feature of this region is high biaxiality of molecular arms.   Though  the biaxial phases  play a dominant role here,  in a small temperature interval, in addition to N$_\text{B}$ and N$_\text{TB,B}$ phases, it is possible to stabilize the  N$_\text{U}$ phase, as well. The last plot (Figure \ref{fig:phase5}) depicts a phase transition between   N$_\text{B}$ and  N$_\text{TB,B}$ at the self--dual point ($\lambda=1/\sqrt{6}$) for the arms  and $\chi=140^\circ$. Here bent--core molecular arms are maximally biaxial (\textit{i.e.} they are neither prolate nor oblate). The N$_\text{B}\leftrightarrow \text{N}_\text{TB,B}$ phase transition is first order as can be deduced from a discontinuity in all order parameters (see Figures \ref{fig:phase4} and \ref{fig:phase5}). 

\section{Summary}
In conclusion, we  analyzed an extension of the generic GLF mean--field model \cite{ref31} to study the role biaxiality of V--shaped molecules can play in the stabilization of $\text{N}_{\text{TB}}$ relative to the nematic and isotropic phases. We  assumed that each of the two arms of a bent--shaped molecule is intrinsically biaxial and took local biaxial \emph{ansatz} for the alignment tensor. In the limit of uniaxial particles ($\chi=0^{\circ}, 180^{\circ}$ and $\lambda=0$)  we  recover mean--field results of Maier and Saupe. For ordinary biaxial molecules ($\chi=0^{\circ}, 180^{\circ}$ and $\lambda\neq0$) the model becomes reduced to mean--field version of the well known Lebwohl--Lasher dispersion model \cite{ref45, ref46}. As all bent--core molecules are biaxial \cite{bcm} our generalization seems important for it allows to control intrinsic molecular biaxiality (by two molecular features: bend angle and arm anisotropy). 

We showed that in our extended model, in addition to  $\text{N}_{\text{U}}$ and $\text{N}_{\text{TB}}$, two extra phases: homogeneous $\text{N}_{\text{B}}$, and periodic $\text{N}_{\text{TB,B}}$ -- the analog of $\text{N}_{\text{TB}}$ with local biaxial order of molecular arms -- can be studied, where $\text{N}_{\text{B}}$ appears in a natural way as a limiting case of $\text{N}_{\text{TB,B}}$. For small bend angles the phase diagram becomes dominated by the $\text{N}_{\text{TB,B}}$ phase with no homogeneous nematics present, even for a relatively small molecular  biaxiality ($\lambda \lesssim 0.18$). Here,  both of the twist--bend structures are reachable directly from the isotropic phase, like in the recently reported experiments \cite{ref14,ref47}. Widening the bend angle opens the path for stabilization of standard nematics, where they start to dominate over less conventional phases as in \cite{ref40,ref41,ref42}.However, the stable $\text{N}_{\text{B}}$ phase is not found for $\chi\lesssim 140^\circ$. 
 
One can see from Figures \ref{fig:phase1}--\ref{fig:phase5} that the asymptotic relation for the tilt angle (\ref{eq:theta_limit}) is actually the limit for $\theta$ in the $\text{N}_{\text{TB,B}}$ phase, as this structures is a ground state for $ 0< \lambda \leq 1/\sqrt{6}$. At the transition between the two twist--bend nematics both the pitch and the cone angle in $\text{N}_{\text{TB,B}}$ are smaller than in the $\text{N}_{\text{TB}}$ phase. 

Finally, the model introduced in this paper can be extended further to include competition between such
molecular/external factors as (steric/electric) dipolar forces and external field(s)\cite{ref36,ref50,ref51}. Then, further
nematic  structures with one--dimensional modulation are also  expected \cite{ref23,ref24,ref25,ref26,ref27,ref22,ref48,ref49}.
\section*{Acknowledgments}
\noindent
This work was supported by Grant No. DEC-2013/11/B/ST3/04247 of the National Science Centre in Poland and in part by PL--Grid Infrastructure. The authors are grateful to Oleg D. Lavrentovich for pointing out the necessity of improvements in Figure \ref{fig:scheme1} and the authors thank Micha\l{} Cie\'sla and Pawe\l{} Karbowniczek for further stylistic comments and suggestions regarding this Figure. The authors would like to thank the Referees for valuable suggestions in improving the manuscript.

\bibliography{rsc} 
\bibliographystyle{rsc} 

\end{document}